\documentclass[preprint, authoryear]{elsarticle}

\usepackage{natbib}
\usepackage[utf8]{inputenc} 
\usepackage[T1]{fontenc}
\usepackage{lmodern}
\usepackage[fleqn]{mathtools}
\usepackage{booktabs}
\usepackage{rotating}
\usepackage{amssymb}
\usepackage{amsmath}
\usepackage{graphicx}
\usepackage{caption}
\usepackage{epstopdf}
\usepackage{stackrel}
\usepackage{soul}
\usepackage{color}
\usepackage{float}
\usepackage[official]{eurosym}
\usepackage{tabularx}
\usepackage{afterpage}
\usepackage{mathrsfs}
\usepackage{chngcntr}
\usepackage{caption}
\usepackage{subcaption}

\renewcommand*{\underline}{\ul}

\newdefinition{rmk}{Remark}
\newproof{pf}{Proof}
\newproof{pot}{Proof of Theorem \ref{thm2}}

\makeatletter
\def\ps@pprintTitle{%
	\let\@oddhead\@empty
	\let\@evenhead\@empty
	\def\@oddfoot{\centerline{\thepage}}%
	\let\@evenfoot\@oddfoot}
\makeatother

\begin{document}
\pagenumbering{gobble}
\title{ Bitcoin: Like a Satellite or Always Hardcore?\\A Core-Satellite Identification in the Cryptocurrency Market
	\\[12pt]
{\small(First Draft: May 14, 2021)}}

\author[hhu]{Christoph J.\ Börner}
\ead{Christoph.Boerner@hhu.de}	
\author[hhu]{Ingo Hoffmann}
\ead{Ingo.Hoffmann@hhu.de}	
\author[hhu]{Jonas Krettek\corref{cor1}}
\ead{Jonas.Krettek@hhu.de}
\author[hhu]{Lars M.\ Kürzinger}
\ead{Lars.Kuerzinger@hhu.de}
\author[hhu]{Tim Schmitz}
\ead{Tim.Schmitz@hhu.de}

\cortext[cor1]{Corresponding author. Tel.: +49 211 81-11515; Fax.: +49 211 81-15316}

\address[hhu]{Financial Services, Faculty of Business Administration and Economics, \\ Heinrich Heine University D\"usseldorf, 40225 D\"usseldorf,
	Germany}	

\begin{abstract}
	Cryptocurrencies (CCs) become more interesting for institutional investors' strategic asset allocation and will be a fixed component of professional portfolios in future. 
	This asset class differs from established assets especially in terms of the severe manifestation of statistical parameters. The question arises whether CCs with similar statistical key figures exist. On this basis, a core market incorporating CCs with comparable properties enables the implementation of a tracking error approach. 
	A prerequisite for this is the segmentation of the CC market into a core and a satellite, the latter comprising the accumulation of the residual CCs remaining in the complement. Using a concrete example, we segment the CC market into these components, based on modern methods from image\;/\;pattern recognition.

\end{abstract}

\begin{keyword}
	Cryptocurrencies 
	\sep Core-Satellite Identification
	\sep Market Segmentation 
	\sep Pattern Recognition					\\[6pt] 
	
	
	\noindent \textit{ORCID IDs:} 
	0000-0001-5722-3086 (Christoph J.~B\"orner), 
	0000-0001-7575-5537 (Ingo Hoffmann), 
	0000-0002-0978-6252 (Jonas Krettek), 
	0000-0001-5774-1983 (Lars M.~K\"urzinger), 
	0000-0001-9002-5129 (Tim Schmitz)			\\[6pt] 
	
	\noindent \textit{Acknowledgement:} We thank Coinmarketcap.com for generously providing the cryptocurrency time series data for our research. 
\end{keyword}
\maketitle
\newpage

\pagenumbering{arabic}
\section{Introduction} \label{introduction}

Cryptocurrencies (CCs) have gained tremendous attention and popularity in media and society in recent years, not least because of the severe manifestation of their statistical parameters, especially market volatility. Due to their nature, CCs are seen more as an investment object than as a currency \citep{Baur.2018} in the classical sense. The development of rising investment volumes has been continuing for years and it can be assumed that CCs are gradually on their way of becoming an established asset class. Against this background, it seems plausible that CCs will become a fixed component of institutional investors' portfolios in the future.\par 
In professional portfolio managagement, one approach is to segment the investment universe into a core of assets with homogenous statistical properties and assets that differ significantly from these properties -- the so-called satellite. The core market can then be tracked using adequate asset picks with a tracking error approach. The satellite investments represent only a small proportion of the total portfolio, which are mostly actively managed sub-portfolios covering selected areas. They are meant to deliver above-average returns and have a diversifying effect due to their low correlation with the core investment \citep{Amenc.2012}.\par
In standard portfolios, for example, satellite investments such as geographical regions, asset classes different from the core investment, and the purchase of portfolios with different management styles or strategies are suitable for enriching or diversifying the core portfolio. It is also possible to consider a certain asset class and differentiate between core investments and satellite investments. A sector selection for corporate bonds or the segmentation of stocks into "with the market" (core) and "high beta stocks" (satellite) serve as an example.\par
In this paper, we are looking specifically at the CC asset class and propose a method to segment the core market from the satellite, based on the development of key statistical parameters.\par 
Attempts to depict the CC market holisticly for reasons of portfolio and risk management have already been investigated in literature. A prominent strand of literature deals with index construction for CCs. In this context, the CC market index CRIX proposed by \cite{Trimborn.2018a} represents a well-known example, which is intended to serve as a starting point to adress these economic questions. A similar top-down approach based on the 30 largest CCs by market capitalization is used to calculate the CC index CCi30 \citep{Rivin.2017}.
\par
However, instead of focusing on market capitalization and trading volume and thus prioritizing larger CCs, we identify the core market by applying a core-satellite approach based on the individual risk-return profile. Our approach has some potential advantages compared to a top-down constructed index. While the indices only take the largest CCs into account and may suffer from survivorship bias, the core-satellite approach identifies the core of the market, i.e.\ those CCs that behave similarly in statistical terms. Even though we currently exlusively consider 27 CCs due to data gaps, in perspective, as the market grows, it might become possible to use our method to identify the core market from a large number of CCs. To build a portfolio, investors would then no longer have to replicate the indices, but could delibaretely buy fewer individual assets of the core of the CC universe and combine them with those of the satellite.
Since the core market could be represented with fewer assets in the portfolio, the monitoring costs for the portfolio would decrease. Moreover, potential problems in the portfolio, such as price collapses, operational risk \citep{Trimborn.2020} or the extinction of entire CCs, could be countered more quickly. This is a decisive advantage, especially due to the dynamics and speed of the market.\par

In order to determine CCs showing a comparable performance over the course of 2014 to 2019, we consider returns as well as standard deviations proposed by modern portfolio theory \citep{Markowitz.1952}. 
One general problem is that CCs are different from traditional asset classes, especially in terms of extreme tails and corresponding tail risk. Against this background, \cite{Majoros.2018} and \cite{Borner.2021} show that the stable distribution (SDI) is well suited to statistically model the returns of CCs overall and especially in the tail area. Thus, we extend our data base by including the tail parameter $\alpha$ of the SDI to specifically consider the tail risk. To identify similarity patterns in the development of statistical parameters, we use the Dynamic Time Warping (DTW) algorithm. This algorithm has originally been developed for speech recognition \citep{Sakoe.1}, but is widely used for clustering and  classification in the fields of various applications today \citep{Giorgino.2009}.\par
The DTW analysis leads to DTW distances that are defined in pairs. The question arises if assets - in the present case CCs - can be grouped together in such a way that they are similar in the sense of short DTW distances according to specified criteria - here the aforementioned statistical indicators. This would allow assets to be divided into a core and a satellite. The particular difficulty lies in the fact that the sorting of the individual DTW distances becomes a monotonically increasing function over the natural numbers and the possible value range $[0, \infty]$ is almost continuously covered in many cases. Based on this, it must be examined whether a specific DTW distance can be derived purely from the data, acting in further steps as a boundary to divide the investment universe into core and satellite. In the following, we present a general procedure that is based on modern methods from pattern recognition and answer precisely these questions. Step by step, we show the separation of core assets within an investment universe. The process is not restricted to a specific asset class and can be used wherever it is important to separate similar from dissimilar assets.\par 
Using the statistical parameters' development, we show that segmenting the CC market into a core and a satellite succeeds when applying our method. Furthermore, we answer the question of whether Bitcoin is indeed part of the hard core of the cryptocurrency market or just a satellite. As the CC market becomes more professional, that is, as market capitalization, liquidity, and market depth increase, the method might become an indispensable tool for professional asset management.
\par
The remainder of this paper is structured as follows: In Sec.\ \ref{dataCC} we describe the data used for our analysis. In Sec.\ \ref{DTW_Methodology} a brief overview of the DTW methodology is given.
In the main part of our study, Sec.\ \ref{CSIdentification}, we develop the identification method to separate the CC universe into a core market segment and a complementary segment which is an accumulation of the residual CCs remaining -- called satellite. The separating procedure is shown using real data. The last section summarizes our most important results and gives an overview of further research topics.

\section{Data} \label{dataCC} 
As a foundation of our analysis, we follow various studies by extracting CCs’ daily prices from the website coinmarketcap.com \citep{Fry.2016, Hayes.2017, Brauneis.2018, Caporale.2018, Gandal.2018, Glas.2019}. In order to depict the CC market as a whole, we aim to include as many CCs in our analysis as possible. However, there is a trade-off between having the longest time series possible and the number of CCs in the sample because, on average, seven CCs per week die out \citep{ElBahrawy.2018}. Against this background, we end up with an observation period from 2014-01-01 to 2019-06-01 taking $66$ potential CCs from the Coinmarketcap Market Cap Ranking at the reference date of 2014-01-01 into consideration, which have been present throughout the entire timeframe.\par 
As data gaps appear in the time series of most CCs, we exclude all CCs with five or more consecutive missing observations. By utilizing the Last Observation Carried Forward (LOCF) approach, as previously done in \cite{Schmitz.2020}, \cite{Trimborn.2020}, \cite{Borner.2021}, we are able to include all CCs with smaller data gaps. Hence, $N = 27$ CCs remain in our data set, as depicted in Tab.\  \ref{tab:DatasetTestAssetNutshell}. 

\begin{table}[H]
	\footnotesize \noindent
	\begin{tabularx}{\textwidth}{  l  X   l  X l  X}
		\toprule
		 CC & ID &  CC & ID & CC & ID\\
		\midrule
		 Anoncoin 		& ANC &
		 BitBar 		& BTB &
		 Bitcoin 		& BTC \\
		 CasinoCoin 	& CSC &
		 Deutsche eMark& DEM &
		 Diamond 		& DMD \\
		 Digitalcoin	& DGC &
		 Dogecoin 		& DOGE& 
		 Feathercoin	& FTC \\
		 FLO 			& FLO &
		 Freicoin 		& FRC & 	 
		 GoldCoin 		& GLC \\ 
		 Infinitecoin 	& IFC &
		 Litecoin  		& LTC &
		 Megacoin 		& MEC \\   
		 Namecoin 		& NMC &
		 Novacoin 	    & NVC &
		 Nxt 			& NXT \\
		 Omni			& OMNI&
		 Peercoin 		& PPC &
		 Primecoin		& XPM \\
		 Quark 			& QRK &
		 Ripple 		& XRP &
		 TagCoin 		& TAG \\
		 Terracoin		& TRC &
		 WorldCoin 		& WDC &
		 Zetacoin 		& ZET  \\

		\bottomrule
	\end{tabularx}
	\caption{: Considered CCs, data source: CoinMarketCap.}
	\label{tab:DatasetTestAssetNutshell}
\end{table}

In a next step, we convert the CC closing prices denoted in USD to EUR prices, using the daily USD-EUR exchange rates retrieved from Thomson Reuters Eikon. To prevent potential weekday biases, the resulting (daily) observations are converted to weekly observations \citep{Dorfleitner.2018, Aslanidis.2020}. The choice of weekly input data derives from the fact that professional asset management by institutional investors often does not operate on the basis of daily data. Intrayday data are not considered to further avoid biases, e.g.\ through pump and dump schemes.\par 
As a starting point, we compute logarithmized weekly returns, which are referred to as returns for the sake of simplicity in the following. Based on this, we calculate the average weekly returns per year as well as the standard deviations and fit the tail parameter $\alpha$ of the SDI.  Our longitudinal analysis from 2014 to 2019 allows to examine the market dynamics of the statistical parameters mentioned.\par
 
\section{Dynamic Time Warping}\label{DTW_Methodology}
As we use three variables simultaneously, a simple correlation analysis is not adequate, since it would only provide information about the similarity of a certain statistical variable and has no valid significance for a $N \times N$-matrix with $N = 27$, with a given total of only six annual measuring points. Furthermore, it does not live up to the dynamic character of the development of the statistical parameters. Therefore, we use the DTW algorithm to segment the market into CCs that show a similiar behaviour over the investigation period.\par
\cite{Sigaki.2019} employ this methodology, revealing clusters of CCs with similar information efficiency. However, they solely consider returns as a variable on a smaller time period. Although pursuing a different goal, we also end up with a DTW distance matrix, but instead of analyzing clusters, we use pattern recognition methods to identify core and satellite CCs. Because the DTW alogorithm is well known and widely used, only a cursory overview over this method is given in the following focusing on relevant features of our analysis. \par 
The DTW distance can be used as a shape-based dissimilarity measure that finds the optimum warping path between two time series by minimizing a cost function \citep{Sakoe.1,Aghabozorgi.2015}. 
By following the definition and notation of the main strand of literature, then in a the first step, a so called distance matrix between each pair of time series compared needs to be calculated. This distance matrix can be based on various metrics. For our analysis and for reasons of robustness, we compute Manhattan, Euclidean and squared Euclidean distance matrices. As explained in Sec.\ \ref{CSIdentification} in more detail, we use three variables per CC to determine the distance matrices between each pair of (multivariate) time series over the course of 2014 to 2019. We end up with a distance matrix for each of the three metrics and each pair of time series. Note, the distance matrix described so far is -- due to the six discrete points in time -- made up of a scheme of six rows and six columns. For a specific currency pair, each cell in the scheme contains the distance in the respective metric for a specific point in time. 
In literature the latter is also referred to as {\it local cost matrix}. 
The above scheme must be carefully distinguished from the distance matrix ${\bf D}$ defined in the following Sec.\;\ref{CSIdentification}.
\par 
Given the distance matrix, i.e.\ the $6\times 6$ scheme of each CC pair, the DTW algorithm finds the optimal alignment through it, starting in each distance matrix at $(2014,2014)$ and finishing at $(2019,2019)$ \citep{Sakoe.1}. It implies that the time differences between the time series are eliminated by warping the time axis of one so that the maximum coincidence is attained with the other \citep{Sakoe.1}. The individual distances of the DTW path are aggregated to total costs using a cost function. The total costs, referred to as DTW distance $d$, reflect the minimum costs between the time series compared. For a better understanding, it should be noted that the DTW distance between the same objects equals 0 since there is no dissimilarity. The upper left part of Fig.~\ref{Fig2CSIdentification} shows the DTW distance $d_{mn}$ for each pair $m,n = 1, \ldots, N$ of the $N=27$ CCs.\par

We outline the underlying methodology only briefly, but there are several restrictions and setting options for the algorithm and the cost function respectively. For a more detailed overview see e.g.\ \cite{Sakoe.1} and \cite{Giorgino.2009}.

\section{Core-Satellite Identification}\label{CSIdentification}
In strategic asset allocation, a core-satellite strategy is the division of the investment into a portfolio consisting of a broadly diversified core investment which is intended to offer a basic return with moderate risk, and several individual investments (satellites) with higher risk and higher earnings potential. The latter serves to increase the return of the overall investment \citep{Methling.2019}.

The returns, sample average $\langle r \rangle$, the standard deviation $s$ and the tail parameters $\alpha$ are examined as essential statistical parameters for CCs. 

A brief overview of the used parameterization and the SDI's main features are given in \ref{Appendix_SDI}. The tail parameter $\alpha$ plays a significant role in differentiating between CCs in which the returns almost obey a normal distribution (i.e.\ $\alpha\rightarrow 2$) or possess a long tail (i.e.\ $\alpha \ll 2$) with correspondingly high tail risks. 

Overall, we consider the dynamics of the sample vector $\left(\langle r \rangle, s, \alpha\right)'$ over time for the years 2014 to 2019 in our analysis.

In Tab.\ \ref{TabelleInputData} an exemplary excerpt of four CCs of the whole data set is shown. The aim is to use the temporal development of the statistical parameters to infer CCs that can be assigned to a market core due to their similar statistical behaviour.

\begin{table}[ht]
	\footnotesize \noindent
	\begin{tabularx}{\textwidth}{llcrr rrrr|r} \toprule								
		\multicolumn{2}{l}{\bf CC}					&	 
		\multicolumn{1}{l}{\bf Value}					&
		\multicolumn{6}{l}{\bf Sample Vector}	&		
		\multicolumn{1}{l}{\bf DTW Dist.}	    	\\ [2pt]
		\multicolumn{1}{l}{No.}									&
		\multicolumn{1}{l}{ID}									&	
		\multicolumn{1}{c}{ }									&		 
		\multicolumn{1}{r}{2014}	&
		\multicolumn{1}{r}{2015}	&
		\multicolumn{1}{r}{2016}	&
		\multicolumn{1}{r}{2017}	&
		\multicolumn{1}{r}{2018}	&
		\multicolumn{1}{r}{2019}	&
		\multicolumn{1}{l}{Metric / $d_{mn}$}					\\	\midrule
		6	&	DMD	&	$\langle r \rangle$	&	-4.52	&	1.67	&	-0.73	&	8.46	&	-5.50	&	0.85	&	\multicolumn{1}{l}{Manh.}	\\
		&		&	$s$	&	28.12	&	22.96	&	9.65	&	20.00	&	14.37	&	16.48	&	6.01	\\
		&		&	$\alpha$	&	1.81	&	1.18	&	2.00	&	2.00	&	1.30	&	2.00	&	\multicolumn{1}{l}{Eucl.}	\\[2.5pt]
		11	&	FRC	&	$\langle r \rangle$	&	-7.10	&	-1.62	&	-0.25	&	5.92	&	-1.97	&	5.05	&	3.10	\\
		&		&	$s$	&	18.74	&	20.27	&	59.19	&	44.84	&	29.43	&	104.04	&	\multicolumn{1}{l}{sq. Eucl.}	\\
		&		&	$\alpha$	&	2.00	&	1.44	&	0.63	&	1.18	&	1.51	&	0.90	&	5.03	\\ \midrule
		21	&	XPM	&	$\langle r \rangle$	&	-7.27	&	-0.38	&	-0.58	&	5.38	&	-2.98	&	0.70	&	\multicolumn{1}{l}{Manh.}	\\
		&		&	$s$	&	15.58	&	24.28	&	8.66	&	26.47	&	22.84	&	13.35	&	1.06	\\
		&		&	$\alpha$	&	1.67	&	1.65	&	1.79	&	1.63	&	1.76	&	1.57	&	\multicolumn{1}{l}{Eucl.}	\\[2.5pt]
		27	&	ZET	&	$\langle r \rangle$	&	-5.74	&	0.16	&	0.51	&	2.88	&	-3.73	&	1.11	&	0.53	\\
		&		&	$s$	&	28.72	&	24.63	&	16.39	&	35.35	&	20.14	&	24.01	&	\multicolumn{1}{l}{sq. Eucl.}	\\
		&		&	$\alpha$	&	1.65	&	1.68	&	1.72	&	1.74	&	1.83	&	1.80	&	0.13	\\
				
		\bottomrule
	\end{tabularx}
	\caption{: Input data for DTW distance analyses (exemplary excerpt). Return $\langle r \rangle$ and standard deviation $s$ in percent per week.}
	\label{TabelleInputData}
\end{table}

The three-dimensional vector $\left(\langle r \rangle, s, \alpha\right)'$ is examined over the course of six years and the DTW distance $d_{mn}$ is determined in pairs, with $m,n = 1, \ldots, N$. The DTW distance is calculated in three different metrics: Manhattan, Euclidean and squared Euclidean. This allows three matrices ${\bf D}_{\text{Metric}}\in \mathbb{R}^{N \times N}$ for the CCs to be determined, each for a specific metric, with elements $d_{mn; \text{Metric}}$. These square matrices are symmetrical $d_{nm} = d_{mn}$, the entries on the diagonal are zero $d_{mm} = 0$ and the off-diagonal elements are all positive.

For two pairs of CCs in the last column of Tab.\ \ref{TabelleInputData} the calculated DTW distances -- in each metric -- can be compared. Note, the number in the first column corresponds to the numbering in Tab.\ \ref{TabelleCoreSatellite}. 

The first pair (Diamond DMD and Freicoin FRC) exemplarily exhibits a considerable distance in each metric.
A detailed analysis of the vectors over time shows the reason for this great distance. On the one hand, clear differences can be observed with regard to the absolute level and the sign (same year) of the returns. On the other hand, there are strong differences in the standard deviation and in the tail parameter (same year). While the standard deviation of the CC Diamond remains almost the same at a high level, the scattering of the returns of the FRC increases dramatically in 2019. For both CCs it is remarkable that the underlying return distribution changes from almost normal to a heavy tail distribution. This can be clearly seen in the change in the tail parameter $\alpha$ (year-to-year). Since this change does not occur at the same time for the two CCs (same year), the DTW analysis results in large distances. Furthermore, we observe this changing distribution behavior with other CCs. At this point, we recommend the fact of a potential time varying, hence, non stationary distribution to be examined more closely in risk controlling of institutional investors if CCs represent a significant component of asset allocation. 

The second pair of CCs (Primecoin XPM and Zetacoin ZET) illustrates two CCs behaving very similarly. Overall, comparatively small DTW distances can be observed here. The returns, the standard deviation and the tail parameters are closely related. It is also noteworthy that the form of the underlying distribution of returns hardly changes; the variability of the tail parameter is likely to derive from statistical errors based on the small database. 

\begin{figure}[htbp]
	\captionsetup{labelfont = bf, labelsep = none}
	\includegraphics[width=0.995\textwidth]{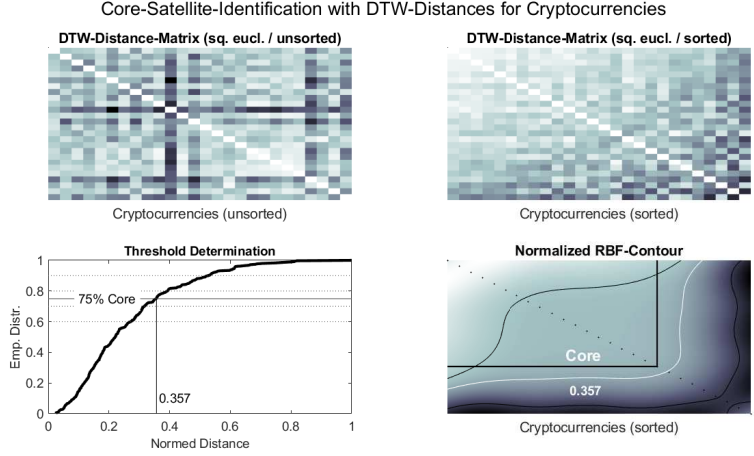}
	\caption[Core-Satellite Identification]{\label{Fig2CSIdentification} Shown is the procedure to identify and separate a core of some CCs within the whole market. The remaining CCs belong to a set which encloses the satellite.}
\end{figure}

The upper left part of Fig.\ \ref{Fig2CSIdentification} shows the distance matrix ${\bf D}_{\text{SE}}$ for the DTW distances in the squared Euclidean metric as a surface plot for all CCs. The ordering descends from the numbers given in the first column of Tab.\ \ref{TabelleCoreSatellite}. The colors used indicate the value of the DTW distance from small (white) to large (black), i.e.\ in this example from 0 to about 5. The entries with zero DTW distance are marked white. 

The problem of identifying groupings in the set of CCs leads to the problem of finding structures in the distance matrix ${\bf D}_{\text{Metric}}$. One possibility to carry out this structural analysis of the distance matrices is to apply methods that have been used for a long time in the investigation of hierarchical matrices \citep{Liu.2012, Hackbusch.2015}. Similar to image recognition, these methods aim to recognize patterns in matrices. 

In a first step, the CCs are rearranged in such a way that the CC displaying the greatest distance to all others on average is depicted on the right. In a descending order, the CCs with the next smaller distances are arranged to the left.\footnote{This form of ordering is the same as sorting according to maximum rows or column total.} As a result we gain an ordered set of CCs and the resulting surface plot changes as shown in the upper right part of Fig.\ \ref{Fig2CSIdentification}.
The similarity of different CCs with regard to the dynamics of the statistical key figures is given when the DTW distance is small and tends towards zero. This is the case for CCs in the upper left white corner in the sorted matrix. Starting from the top left corner in the direction of the main diagonal up to a certain distance $d_{\text{bound}}$, the CCs thus delimited would represent the market core of similar CCs.

Since the height profile above the sorted distance matrix has a peaked, rough structure, cf.\ Fig.\ B.2 in the appendix, the delimitation of the set of CCs belonging to the core cannot be carried out reliably. Therefore, a modeling is first carried out, which represents the height profile more smoothly. We use a modeling method comparable to the analysis of hierarchical matrices, cf.\ e.g.\ \cite{Hackbusch.2015} and the corresponding literature cited therein. 

There is a certain basic structure of the matrix that simplifies the modeling problem. The sorted distance matrix is square, symmetrical and has only positive elements. To the right and lower edge of the sorted distance matrix, the entries become larger on average, so that a concave structure is essentially present. In addition, we are only looking for a certain block in the sorted matrix, which starts in the upper left corner and is itself square.

The surface's concave structure can be modeled well with radial basis functions, which have their centring points -- similar to a frame -- in the outer area of the edges of the sorted distance matrix. A brief overview of the radial basis function model class used is given in \ref{Appendix_RBF}. If the individual elements of the distance matrix are normalized between 0 and 1, the modeling leads to an area whose height profile can be seen in the lower right part of Fig.\ \ref{Fig2CSIdentification}.

In a next step, we define the boundary condition $d_{\text{bound}}$, which delimits the set of CCs belonging to the core. The distance matrix solely incorporates positive elements, but it is not positive definite in all cases so that some eigenvalues can be negative and an analysis of the eigenvalue spectrum does not lead to the definition of a suitable threshold $d_{\text{bound}}$. 

Instead of this consideration, we analyze the empirical distribution function over the elements of the upper triangular matrix normalized between 0 and 1 and delimit an area that contains $p$-percent of the smallest distances, cf.\ the lower left part of Fig.\ \ref{Fig2CSIdentification}. In practice the $p$-value will be somewhere between 
$60\% \ldots 90 \%$ (dotted lines), where the steep slope of the empirical distribution function merges into the flatter area. This represents the tail area of the empirical distribution function in which the empirically measured DTW distances increase rapidly. 
If small values for $d_{\text{bound}}$ are found, the associated core is more homogeneous. The larger, the more heterogeneous the core becomes with regard to the statistical parameters. 
In our example we find the kink at about $p \approx 75\%$. The associated threshold $d_{\text{bound}} = 0.357$ is used to draw a contour (white) in the modelled height profile on the right, which delimits an upper block matrix.

This block matrix describes the market core if the squared Euclidean metric is used. Table \ref{TabelleCoreSatellite}, penultimate column, shows the CCs belonging to the core if this metric is utilized (boolean 1 indicates: belonging to the core). 

\begin{table}[ht]
	\footnotesize \noindent
	\begin{tabularx}{\textwidth}{lllcccc} \toprule								
		\multicolumn{2}{l}{\bf CC}					&	 
		\multicolumn{1}{l}{\bf Name}					&
		\multicolumn{3}{l}{\bf Metric}					&		
		\multicolumn{1}{l}{\bf Core}	    			\\  [2pt]
		\multicolumn{1}{l}{No.}							&
		\multicolumn{1}{l}{ID}							&	
		\multicolumn{1}{c}{ }							&		 
		\multicolumn{1}{r}{Manhattan}					&
		\multicolumn{1}{r}{Euclidean}					&
		\multicolumn{1}{r}{Sq. Eucl.}					&
		\multicolumn{1}{l}{Intersection}				\\	[2pt]
		\multicolumn{2}{l}{ }							&
		\multicolumn{1}{r}{\scriptsize Threshold:}    				&	
		\multicolumn{1}{c}{\scriptsize 0.554}						&		 
		\multicolumn{1}{c}{\scriptsize 0.564}						&
		\multicolumn{1}{c}{\scriptsize 0.357}						&
		\multicolumn{1}{l}{ }							\\	\midrule
		1	&	ANC	&	Anoncoin	&	1	&	1	&	1	&	{\bf C}	\\
		2	&	BTB	&	BitBar	&	1	&	1	&	1	&	{\bf C}	\\
		3	&	BTC	&	Bitcoin	&	1	&	1	&	1	&	{\bf C}	\\
		4	&	CSC	&	CasinoCoin	&	0	&	0	&	0	&	S	\\
		5	&	DEM	&	Deutsche.eMark	&	1	&	1	&	1	&	{\bf C}	\\
		6	&	DMD	&	Diamond	&	0	&	1	&	0	&	S	\\
		7	&	DGC	&	Digitalcoin	&	1	&	1	&	1	&	{\bf C}	\\
		8	&	DOGE	&	Dogecoin	&	1	&	1	&	0	&	S	\\
		9	&	FTC	&	Feathercoin	&	1	&	1	&	1	&	{\bf C}	\\
		10	&	FLO	&	FLO	&	1	&	1	&	1	&	{\bf C}	\\
		11	&	FRC	&	Freicoin	&	0	&	0	&	0	&	S	\\
		12	&	GLC	&	GoldCoin	&	1	&	1	&	1	&	{\bf C}	\\
		13	&	IFC	&	Infinitecoin	&	0	&	0	&	0	&	S	\\
		14	&	LTC	&	Litecoin	&	1	&	1	&	1	&	{\bf C}	\\
		15	&	MEC	&	Megacoin	&	1	&	1	&	1	&	{\bf C}	\\
		16	&	NMC	&	Namecoin	&	1	&	1	&	1	&	{\bf C}	\\
		17	&	NVC	&	Novacoin	&	1	&	1	&	1	&	{\bf C}	\\
		18	&	NXT	&	Nxt	&	1	&	1	&	1	&	{\bf C}	\\
		19	&	OMNI	&	Omni	&	1	&	1	&	1	&	{\bf C}	\\
		20	&	PPC	&	Peercoin	&	1	&	1	&	1	&	{\bf C}	\\
		21	&	XPM	&	Primecoin	&	1	&	1	&	1	&	{\bf C}	\\
		22	&	QRK	&	Quark	&	1	&	1	&	1	&	{\bf C}	\\
		23	&	XRP	&	Ripple	&	0	&	0	&	0	&	S	\\
		24	&	TAG	&	TagCoin	&	1	&	1	&	0	&	S	\\
		25	&	TRC	&	Terracoin	&	1	&	1	&	1	&	{\bf C}	\\
		26	&	WDC	&	WorldCoin	&	0	&	0	&	0	&	S	\\
		27	&	ZET	&	Zetacoin	&	1	&	1	&	1	&	{\bf C}	\\
		\bottomrule
	\end{tabularx}
	\caption{: Analysis of the DTW distance for the different metrics. Identification of the core (C) and the satellite set (S).}
	\label{TabelleCoreSatellite}
\end{table}

In our analysis, we examine the DTW matrices for all metrics in the same way. The CCs belonging to the core according to the respective metric are shown in Tab.\ \ref{TabelleCoreSatellite}.
Note that this method can also be used to find very similar CCs with almost the same statistical behavior if the lower kink is identified in the empirical distribution function and the bound $d_{\text{bound}}$ is thus determined. We have also examined this path of segmentation (not explicitly shown her). This segmentation leads to the delimination of five CCs in the white upper left corner of the lower right part in Fig.\ \ref{Fig2CSIdentification}, which behave almost identically considering the statistical key figures over a long period of time.\par
In Tab.\ \ref{TabelleCoreSatellite} it can be seen that the amount of CCs belonging to the core depends on the metric. In portfolio management it might come down to a decision of practicability which metric to use and which dependency to accept. However, this dependency can be avoided by considering all metrics and selecting those CCs as the market core, which are contained in the intersection of all metrics. This approach is illustrated in the last column of Tab.\ \ref{TabelleCoreSatellite}. In this column, all CCs belonging to the core according to all metrics are marked with {\bf C}. This knowledge provides a decisive advantage in asset management, when integrating a certain share of CCs in a portfolio.\par Table \ref{TabelleInputData} shows examples of CCs as representatives of the identified core (Primecoin XPM and Zetacoin ZET) and the identified satellite (Diamond DMD and Freicoin FRC). In practice, a simple portfolio could be constructed as follows. For example, 5 - 10 CCs with high liquidity and market depth, which are similar to XPM/ZET and belong to the core, are selected from the entire data set. These CCs form the core investment. Individual CCs can then be selected from the satellite, which can be expected to offer a higher return if the risk is higher. In the first case, the tracking error can be determined in relation to the core and in the second case in relation to the overall market. In any case, the composition is optimized taking a specified limitation of the tracking error into account. Continuous control of the tracking error and tactical readjustment of the weights leads to tracking of the core (first case) or the overall market (second case), whereby the tracking error specified by the institutional investor is adhered to.
\section{Conclusion} 
In our study, we show how a general, purely data-driven process can be set up successfully to separate an investment universe into similar assets (core) and dissimilar assets (satellite). We prove the feasibility of this approach and outline the necessary sequence of steps for the segmentation in detail. Using the example of the modern CC asset class, we carry out the separation of the investment universe into similar CCs (core) and dissimilar CCs (satellites) as the residual share. In addition, we ascertain interesting results concerning specifically the CCs.\par
The question raised at the beginning of whether Bitcoin actually represents the hard core of the cryptocurrency market can be answered in a differentiated manner. It turns out that although Bitcoin is counted as part of the core, it rather marks the edge of the core affiliation. A dominant role, which appears in other analyses, cannot be confirmed.\par
Our proposed segmentation can be used in portfolio management by institutional investors to track the core market with a few selected CCs in a tracking error approach. In order to increase returns, a higher-level management approach can then be used to build up individual positions in CCs that belong to the satellite, thus implementing a core-satellite portfolio.\par 
One potential challenge for this approach might lie in liquidity problems, especially in the case of smaller altcoins (other than Bitcoin). However, studies indicate CCs to make up a smaller component of a portfolio of traditional assets, mitigating this issue \citep{Dorfleitner.2018, Schmitz.2020}. In addition, methods such as the Liquidity Bounded Risk-return Optimization (LIBRO) approach by \cite{Trimborn.2020} exist, which can be used to perform portfolio optimization under liquidity constraints. Furthermore, it is conceivable that liquid CCs are incorporated in the core so that they can be purchased anyway without the fear of liquidity restrictions. Beyond that, it can be assumed that the development of the CC market will make it suitable for larger investment volumes in the future.\par
As already mentioned, the proposed method is not limited to CCs. A suitable market segmentation in other asset classes is conceivable, as well. The advantages of product-based implementation of a topic-centered, combined 'core-satellite \& tracking-error' strategy in the private or institutional investor segment, is reserved for further studies. 
\newpage
\appendix
\section{Stable distribution -- the tail parameter $\alpha$}\label{Appendix_SDI}
The analyses in \cite{Borner.2021} showed that the family of SDIs is the most promising for modeling the distribution of the returns of the CCs. 
Therefore, this family of functions is also used in the present study and will be introduced here in detail. Several different parametrizations exist for the SDI. In the following formulation we follow the presentation and the parametrization of the SDI described in \citet[Def.\ 1.4 therein]{Nolan.2020}.

SDIs are a class of probability distributions suitable for modeling heavy tails and skewness. A linear combination of two independent, identically distributed stable distributed random variables has the same distribution as the individual variables.
A random variable $X$ has the SDI $S(\alpha, \beta, \gamma, \delta)$
if its characteristic function is given by:
\begin{flalign}\label{SDI}\nonumber
& \text{E}\left[ \exp\left(\text{i}tX\right) \right]   = \\[8pt]
&
\begin{cases}
\exp\left(\text{i} \delta t - \left|\gamma t\right|^{\alpha}\Big[
1 + \text{i} \beta\text{sign}(t)\;\tan\left(\frac{\pi\alpha}{2}\right) \left(\left|  \gamma t\right|^{1-\alpha} - 1 \right)
\Big] \right)
& \alpha \neq 1 \\[8pt]
\exp\left(\text{i} \delta t - \left|\gamma t\right|\hphantom{{^{\alpha}}}\Big[
1 + \text{i} \beta\text{sign}(t)\;\frac{2}{\pi}
\ln\left(\left|  \gamma t\right| \right)
\Big]  \right) 
& \alpha = 1 
\end{cases}
\end{flalign}

The first parameter $0<\alpha\leq 2$ is called the shape parameter and describes the tail of the distribution. Sometimes this parameter is also denoted as a {\it tail parameter}, {\it index of stability} or as {\it characteristic exponent}.
The second parameter $-1\leq \beta \leq +1$ is a skewness parameter. If $\beta = 0$, then the distribution is symmetric otherwise left-skewed ($\beta<0$) or right-skewed ($\beta>0$). When $\alpha$ is small, the skewness of $\beta$ is significant. As $\alpha$ increases, the effect of $\beta$ decreases. Further, $\gamma\in\mathbb{R^+}$ is called the scale parameter and $\delta\in\mathbb{R}$ is the location parameter.

For the special case $\alpha = 2$ the characteristic function Eq.\ (\ref{SDI}) reduces to 
$\text{E}\left[ \exp\left(\text{i}tX\right) \right] = \exp\left(\text{i} \delta t - (\gamma t)^{2}\right)$ 
and becomes independent of the skewness parameter $\beta$ and
the SDI is equal to a normal distribution with mean $\delta$ and standard deviation $\sigma = \sqrt{2}\gamma$.
This is an important property for portfolio theory, for example, when considering multivariate distributions. Because it is basically possible to model normally distributed components of a random vector with the same function class.

In the main part the tail parameter $\alpha$ is estimated for each year under consideration on a weekly return basis and used as input data for the DTW distance analyse.

\section{Modeling with radial basis functions}\label{Appendix_RBF}
In many scientific areas \citep{Powell.1977, Poggio.1990, Sahin.1997, Biancolini.2017}, radial basis functions are used to carry out a function approximation of the following form:
\begin{flalign}\label{RBFModel}
y({\bf x}) & = \sum_{m=1}^{M} \lambda_m \,\phi(\| {\bf x} - {\bf x}_m \|)
\end{flalign}
where $y({\bf x})$ is a one dimensional function depending on ${\bf x} \in \mathbb{R}^n$. 
The function $y({\bf x})$ is modelled as a sum of $M$ radial basis functions, each centred at a different centre ${\bf x}_m$, and weighted with an appropriate coefficient $\lambda_m$.
The real value of every radial basis function is strictly positive and only depends on the distance between the point ${\bf x}$ and the centre ${\bf x}_m$. The distance $r = \| {\bf x} - {\bf x}_m \| $ is determined in a previously defined norm. We only use the Euclidean distance as the norm in our analyses. 

To model and reconstruct the height profile over the distance matrix ${\bf D}$ in Sec.\ \ref{CSIdentification}, we use radial basis functions of the Gauss type 
\begin{flalign}\label{RBFGausskern}
\phi(r) & = \exp\left(-ar^2\right)
\end{flalign}

with infinite support and a positive shape parameter $a$. The latter can also be interpreted as the effective range of the radial basis function. If $R$ denotes the distance between two different centres and $0<p<1$ denotes the desired residual effect at the next centre, then the area of effect can be set by $a$ due to: $a = - R^{-2}\ln p$.

The parameter vector $\boldsymbol{\lambda}$ is determined using a least square approach. In some applications we found that the least squares fit had problems with ill-conditioned matrices. Therefore we extend our Lagrange function to be minimized by a regularization term. The latter term is also referred to as cost-functional and takes into account the costs of the deviation from a smooth function. The theoretical foundations of this approach goes back to early work from \citet{Tikhonov.1943, Tikhonov.1963}. The implemented regularization procedure is nowadays standard \citep{Poggio.1990}, cf.\ also \citet{Sahin.1997, Biancolini.2017} and the huge amount of literature cited therein. 

Hence, we set the Lagrange function
\begin{flalign}\label{RBFLagrange}
{\cal L} & = \sigma^2 + \alpha \boldsymbol{\lambda}'\boldsymbol{\lambda}
\end{flalign}
with $\sigma^2$ is the squared error between modelled, $\hat y_i$, and sample values, $y_i$, for $i = 1, \ldots, N$ with sample length $N$. Furthermore, $\alpha$ is a positive real number called the regularization parameter. If  $\alpha\rightarrow 0$, the problem is unconstrained and the resulting model can be completely determined from the sample. 
On the other hand, if, $\alpha\rightarrow \infty$, the a priori desired smoothness of the resulting model dominates and leads to a highly smooth function, in the limit nearly flat and almost independent of the measured sample.

Finally, the solution to the minimization problem Eq.\ \ref{RBFLagrange} is
\begin{flalign}\label{RBFSolution}
\boldsymbol{\lambda} & = 
\left( \boldsymbol{\Phi} + \frac{\alpha}{N} {\bf E}\right)^{-1} {\bf v}.
\end{flalign}
Abbreviate $\phi_{im} = \phi(\| {\bf x}_i - {\bf x}_m \|)$ as the value of the $m$-th radial basis function at the sample point ${\bf x}_i$ for $i = 1, \ldots, N$ and given the output $y_i$, than vector 
${\bf v}= {\left( \left\langle  y_i \phi_{im} \right\rangle\right)}_{m = 1, \ldots, M}$ 
and matrix 
$\boldsymbol{\Phi} = {\left( \left\langle \phi_{ik} \phi_{im} \right\rangle \right)}_{k, m = 1, \ldots, M}$ with $\langle \cdot \rangle$ denotes the sample average. Further, ${\bf E}$ denotes the identity matrix in $\mathbb{R}^{M\times M}$.

In practice in very few applications we assigned successive increasing values $0 \leq\alpha< 100$ to the regularization parameter until the observable local roughness or heavily peaked structure of the modelled surface vanishes. We observed that the height profile of the distance matrix ${\bf D}$ is still well reconstructed, but the absolute height was modelled worse with increasing influence of the regularization.
The modelling properties improve if a constant term is additively added to the model Eq.\ (\ref{RBFModel}). The solution Eq.\ (\ref{RBFSolution}) do not change if only the number of radial basis functions is increased by 1, $M \rightarrow M+1$, and the value identical to 1 is assigned to the first radial basis function, $\phi_{1} = 1$ for all ${\bf x} \in \mathbb{R}^n$, and the changes are considered in the elements of vector ${\bf v}$ and Matrix $\boldsymbol{\Phi}$.

The majority of the analyses could be carried out with $\alpha = 0$ and led without regularization procedures to very good results. For the results shown in the main part we have no regularization procedure applied.

\begin{figure}[htbp]   
	\captionsetup{labelfont = bf, labelsep = none}
	\includegraphics[width=0.995\textwidth]{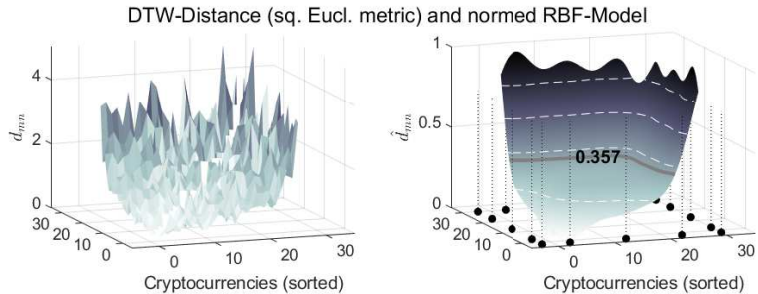}
	\caption[RBF model]{\label{Appendix_DTWDistanzRBFModell} An explicit and detailed representation of the modeling process that led to the segmentation of the CC market in Sec.\ \ref{CSIdentification}. }
\end{figure}

In Fig.\ \ref{Appendix_DTWDistanzRBFModell} an example of the modelling process with radial basis functions and $\alpha = 0$ is shown.
The left picture shows the rough and peaked height structure $d_{mn}$ above the DTW distance matrix ${\bf D}_{\text{SE}}$ calculated in Sec.\ \ref{CSIdentification}. It is the 3 dimensional counterpart of the upper right panel of Fig.\ \ref{Fig2CSIdentification} viewed from the upper left corner along the main diagonal. The graphic on the right shows the surface of the standardized height structure ${\hat d}_{mn}$ modeled with radial basis functions. Also shown some contour lines (dashed white), each with a distance of 0.2 units. The contour to threshold $d_{\text{bound}} = 0.357$ for the squared Euclidean metric is shown in light grey, cf.\ Sec.\ \ref{CSIdentification}.
The bullets and the corresponding vertical dashed lines illustrate the centers and the respective position of the radial basis functions. 
\newpage


\bibliography{../../../FP_03_FatTailsCC/060_Publikationen/01_Tails/010_CryptoTailPaper_v01/LitCC_BundT}

\end{document}